\documentclass[fleqn,usenatbib]{mnras}

\usepackage{newtxtext,newtxmath,amsmath}
 
\usepackage[T1]{fontenc}

\DeclareRobustCommand{\VAN}[3]{#2}
\let\VANthebibliography\thebibliography
\def\thebibliography{\DeclareRobustCommand{\VAN}[3]{##3}\VANthebibliography}

\usepackage{graphicx}	
\usepackage{amsmath}	
\usepackage{amssymb}	
\usepackage{tabularx}
\usepackage{multirow}
\usepackage{makecell}
\usepackage{mathtools}
\usepackage{amsmath}
\usepackage{bm}
\usepackage{mathrsfs}

\title[Visibility plane mosaicing with MeerKAT]{A direction-dependent framework for visibility plane mosaicing and primary beam correction}

\author[K. S. Trehaeven et al.]{K. S. Trehaeven$^{1,2}$\thanks{E-mail: ktrehaeven@gmail.com},
        C. Tasse$^{3,1}$
        O. Smirnov$^{1,4,2}$,
        T. Venturi$^{2,1}$
	\\
    \\
	$^{1}$Centre for Radio Astronomy Techniques and Technologies (RATT), Department of Physics and Electronics, Rhodes University, Makhanda 6140, South Africa\\
	$^{2}$INAF – Istituto di Radioastronomia, via P. Gobetti 101, 40129 Bologna, Italy \\
        $^{3}$GEPI \& USN, Observatoire de Paris, CNRS, Université Paris Diderot, 5 place Jules Janssen, 92190 Meudon, France\\
	$^{4}$South African Radio Astronomy Observatory, Cape Town 7700, South Africa
}

\date{Accepted 2025 December 17. Received 2025 December 07; in original form 2025 August 05}

\pubyear{...}

\begin{document}
	\label{firstpage}
	\pagerange{\pageref{firstpage}--\pageref{lastpage}}
	\maketitle
	
	\begin{abstract}
        With the increasing sensitivity of modern radio interferometers, it has become important to image objects larger than the field of view while optimising sensitivity and image fidelity. We present a coherent visibility plane direction-dependent imaging, calibration and mosaicing framework. Our simulations and application to real MeerKAT data show that this joint deconvolution and primary beam correction approach, coupled with direction-dependent calibration, allows for deeper mosaics with greater fidelity and increased accuracy of recovered flux densities and spectral indices, especially beyond the half-power beam width. Our best-case mosaic produces precise flux values within a 6\% uncertainty and spectral indices within 20\% throughout the imaged area, and is fully complete out to twice the radii and half the flux density than the image plane equivalent. The application to archival wideband MeerKAT 1283 MHz data produces the deepest high-resolution image of the Shapley Supercluster Core, with a sensitivity of 3.6 $\mu$Jy/beam within the primary beam at a 7$^{\prime\prime}$ resolution, constituting a $\sim$50\% increase in dynamic range over the image plane counterpart, and a fluxscale that is consistent within 10\% across the entire field of view. The compute time for the direction-dependent visibility plane mosaic was comparable to the sum of the times needed to perform direction-dependent calibration on the individual pointings. Our results suggest that visibility plane mosaicing with its capability for deeper deconvolution could improve the efficiency of deep and wide surveys, particularly for on-the-fly mapping and studies of low surface brightness sources, and could form the basis of future calibration pipelines for SKA-scale instruments. 
	\end{abstract}
	
	
	
	
	\section{Introduction}



        The current generation of radio interferometers offers unprecedented capabilities in sensitivity, bandwidth, and angular resolution. Such instruments like the Australian Square Kilometre Array Pathfinder \citep[ASKAP,][]{2021PASA...38....9H}, the LOw Frequency ARray \citep[LOFAR,][]{2013A&A...556A...2V}, the Meer Karoo Array Telescope \citep[MeerKAT,][]{2016mks..confE...1J}, and the Upgraded Giant Metrewave Radio Telescope \citep[uGMRT,][]{2017CSci..113..707G} are driving a new era of large-scale radio surveys that aim to uncover faint, diffuse emission and trace cosmic magnetism, galaxy evolution, and the baryon cycle at high fidelity \citep{2012ARA&A..50..455F,2013PASA...30...20N,2016mks..confE...6J,2017A&A...602A...2S,2017PASA...34...52M,2017ARA&A..55..389T,Blyth:2018RG,2019SSRv..215...16V}. With the Square Kilometre Array (SKA) now under construction, these challenges are set to intensify, with enormous data volumes and dynamic range requirements that will push imaging pipelines to their limits. However, scalable and reproducible workflows are required to fully exploit the era of "Big Data" in radio astronomy with tools like parallel processing, high performance computing and cloud infrastructure such as that implemented in the \textsc{Codex Africanus} ecosystem \citep{2024arXiv241212052P,2025A&C....5200962K,2024arXiv241210073B,2025A&C....5200959S}. 

        In particular, a growing body of science now depends critically on the ability to detect and characterise extended, low surface brightness emission across large fields of view. Such emission is often diffuse, filamentary, and spectrally steep, and includes a range of physical phenomena: radio halos and relics in merging clusters \citep{2014IJMPD..2330007B,2019SSRv..215...16V}, intercluster radio bridges \citep{2019Sci...364..981G,2020MNRAS.499L..11B,2022A&A...660A..81V}, and fossil plasma emission from dying Active Galaxtic Nuclei (AGN) lobes or phoenix sources \citep{2015MNRAS.448.2197D,2024ApJ...975..125R,2025A&A...696A.239B}. 
        
        One of the most demanding aspects of modern widefield radio imaging is the accurate recovery of flux density and spectral properties of such sources, which is essential for tracing the life cycle of cosmic rays and the role of turbulence and shocks in particle (re)-acceleration processes \citep{2023A&A...672A..43C,2023A&A...680A..30C,2024A&A...689A.218P}. These goals are often hindered by baseline-time-frequency direction-dependent effects (DDEs) and associated calibration artefacts \citep{2011A&A...527A.106S,2011A&A...527A.107S,2011A&A...527A.108S,2017AJ....154...56J,2018AJ....155....3J,2022AJ....163...87S}. Such effects, which can significantly compromise image fidelity, include complex primary beam (PB) patterns resulting from pointing errors, dish deformation and heterogeneous arrays, \citep{2008A&A...487..419B,2023AJ....165...78D} and associated beam squint and rotation, as well as direction-dependent phase shifts and scintillative decoherence resulting from ionospheric distortions and its associated Faraday rotation \citep{2013A&A...552A..58S,2016RaSc...51..927M,2019MNRAS.483.4100P,2023MNRAS.519.5723O}. Reaching the thermal noise and thereby maximising the dynamic range of deep radio surveys typically necessitates direction-dependent calibration (also called third-generation calibration, or 3GC), with the DDEs explicitly modelled and incorporated into the imaging and deconvolution pipelines, for example, the TIFR GMRT Sky Survey \citep[TGSS,][]{2017A&A...598A..78I}, the LOFAR Two-meter Sky Survey \citep[LoTSS,][]{2021A&A...648A...1T} and the MeerKAT International Gigahertz Tiered Extragalactic Explorations \citep[MIGHTEE,][]{2022MNRAS.509.2150H}.
        
        There are several ways to approach the DD problem in calibration and imaging. The non-coplanarity of modern arrays and their complex primary beams can be corrected during imaging using AW-projection \citep{2008A&A...487..419B,2013A&A...553A.105T}, image-domain gridding \citep[IDG,][]{2018A&A...616A..27V} or faceting \citep{2016ApJS..223....2V,2016MNRAS.460.2385W}. The effects of the ionosphere can be mitigated by peeling individual sources of contaminating artefacts \citep{2018MNRAS.478.2399K}, which involves solving for an additional gains term in the direction of the source and subtracting its contribution from the original visibilities. Alternatively, the DDEs can be modelled directly using Wirtinger calculus and complex-valued optimisation first introduced by \citet{Wirtinger1927}, and used in \textsc{killMS} \citep{2014A&A...566A.127T,2015MNRAS.449.2668S}. Which approach to use, or combination of approaches, depends on the particular science goals and user requirements. For example, AW-projection provides a smooth implementation of DDEs and is hence more accurate, but is difficult to model and becomes computationally expensive. On the other hand, facetting provides a piecewise constant implementation and is therefore faster but loses some accuracy. In reality, W-projection is commonly used in CASA's \textsc{tclean} \cite{2022PASP..134k4501C} and in \textsc{WSClean} \citep[][or rather w-stacking in this case]{offringa-wsclean-2014}, and facetting in \textsc{DDFacet} \citep{2018A&A...611A..87T}, with the exact primary beam implementation depending on the availability of appropriate models for each instrument. For example, in the case of MeerKAT, \textsc{DDFacet} can apply holographic models \citep[][optionally per station]{2023AJ....165...78D} during deconvolution, while \textsc{WSClean} currently applies a symmetric analytic approximation \citep{2020ApJ...888...61M} in the image domain after deconvolution.  

        Future surveys aim to produce images at arcsecond-scale resolution and $\mu$Jy sensitivity across thousands of square degrees \citep{2015aska.confE..67P,2018ASPC..517....3M,2021MNRAS.500.3821B}. The data volumes involved, along with the wide range of spatial and spectral scales present in each pointing, demand a shift away from basic image plane pipelines and toward joint, visibility plane calibration and imaging. These strategies provide a natural framework for incorporating 3GC and enable calibration against deeper and more representative skymodels, improving both fidelity and automation potential for next-generation surveys.

        Traditional mosaicing approaches typically involve stitching together individually imaged and PB corrected fields in the image plane, weighted by the variance of each power beam \citep{1996A&AS..120..375S,2013ApJ...770...91B}. Common software used for such mosaicing are \textsc{MONTAGE}\footnote{\url{http://montage.ipac.caltech.edu/}} \citep{2017PASP..129e8006B} and the \textsc{linearmosaic} CASA tool. However, such methods suffer from limitations including the inability to accurately correct for time-variable beam rotation, particularly at the edges of the PB where models are uncertain \citep{2009IEEEP..97.1472R}. In contrast, mosaicing in the visibility plane has two key advantages: (1) PB correction can be naturally incorporated during deconvolution, and (2) the cumulative sensitivity allows for fainter sources to be deconvolved so that calibration can be performed against a deeper joint skymodel. This leads to more accurate reconstructions, especially when combined with 3GC. \citet{2016AJ....152..124R} compared various wideband imaging methods to determine the accuracy with which intensities and spectral indices can be recovered and found that a joint mosaicing approach, i.e., in the visibility plane before deconvolution, reduced systematic errors compared to stitched mosaics, i.e., in the image plane after deconvolution. We extend this result and incorporate DDEs into the joint mosaic.
        
        In this work, we present a fully direction-dependent framework for mosaicing, and present the first 3GC visibility plane mosaic and PB corrected image using the MeerKAT interferometer \citep{2016mks..confE...1J}. Using both simulations and real MeerKAT data of the Shapley Supercluster, we demonstrate that visibility plane mosaicing with the facet-based imager \textsc{DDFacet}\footnote{\url{https://github.com/saopicc/DDFacet}}, coupled with 3GC using its accompanying solver \textsc{killMS}\footnote{\url{https://github.com/saopicc/killMS}}, substantially improves image fidelity, flux recovery, and spectral index accuracy over the traditional image plane method. Our results suggest that this approach is highly beneficial for deep surveys and diffuse source studies and could form the basis of future calibration pipelines for MeerKAT and SKA-scale instruments.

        This paper is structured as follows: Section 2 reviews the theory of visibility plane mosaicing. In Section 3, presents a suite of simulations testing this theory in a direction-dependent framework. In Section 4, we describe our test field, the Shapley Supercluster and apply our method to real MeerKAT data to showcase the capabilities of visibility plane mosaicing in a practical setting. In Section 5, we discuss the implications of our results, and Section 6 summarises and concludes the work. 
        
        Throughout the paper we assume a $\Lambda$CDM cosmology, with H$_0$ = 70 km/s and $\Omega_{\rm M}$ = 0.3. We adopt the spectral index convention of S $\propto \nu^{-\alpha}$. We use the current (as of writing) stable master branches of the following software \textsc{CARACal v1.1.2}, \textsc{Stimela2 v2.0rc2}, and compatible versions of \textsc{DDFacet} and \textsc{killMS} coming in the next public release.

        \section{Visibiltiy plane mosaicing}

        Radio interferometers form images of the sky by sampling the Fourier transform of the brightness distribution through measurements of the complex visibility function \citep{1999ASPC..180.....T}. Since visibilities are only sparsely measured at discrete points in the spatial frequency space, they must be interpolated (or gridded) onto a regular grid before applying an inverse Fourier transform to obtain the dirty image \citep{1999ASPC..180.....T}. Imaging objects larger or beyond the primary beam of a single pointing requires mosaicing, i.e. combining observations from multiple pointings to reconstruct a wide-field image with uniform sensitivity. Traditional mosaicing methods operated in the image domain \citep{1999ASPC..180..151C}, but visibility domain approaches better handle DDEs, such as the antenna primary beam response, and preserve accurate noise properties across the mosaic \citep{2016AJ....152..124R,Bhatnagar2021ALMA,2022AJ....163...87S}. A key insight enabling visibility plane mosaicing is that the phase centre of an observation need not coincide with the pointing centre: the pointing centre defines the direction of the antenna primary beam, while the phase centre sets the reference direction where the interferometer phases are zero \citep{1999ASPC..180.....T}. By applying appropriate phase rotations, visibilities from different pointings can be shifted to a common phase centre and coherently combined, after which imaging proceeds on a single tangent plane.

        
        Using the framework established by \citet{2018A&A...611A..87T}, the forward (image to visibility) linear transformation of the sky signal to measured visibilities for a single baseline-time-frequency sample is given by their Equation 18:
\begin{equation}
    \mathbf{v}_{\mathbf{b}_{\nu}} = \mathbf{S}_{\mathbf{b}_{\nu}} \mathbf{\mathcal{F}} \mathbf{\mathcal{M}}_{\mathbf{b}_{\nu}} \mathbf{x}_{\nu} \stackrel{\mathrm{def}\mathbf{\mathcal{A}}_{\mathbf{b}_{\nu}}}{=} \mathbf{\mathcal{A}}_{\mathbf{b}_{\nu}} \mathbf{x}_{\nu}.
    \label{eq:1}
\end{equation}
       Here, $\mathbf{v}_{\mathbf{b}_{\nu}}$ denotes the Stokes visibility 4-vector sampled by baseline $\mathbf{b}$ between antenna $p$ and $q$ at time $t$ and frequency $\nu$, $\mathbf{S}_{\mathbf{b}_{\nu}}$ is the sampling matrix, $\mathbf{\mathcal{F}}$ the Fourier transform matrix, $\mathbf{\mathcal{M}}_{\mathbf{b}_{\nu}}$ a Mueller-like matrix encoding the DDEs for this baseline-time-frequency sample, and $\mathbf{x}_{\nu}$ is the true sky brightness distribution vector at frequency $\nu$. This mapping is stacked per visibility to produce the full mapping function. In the case of visibility plane mosaicing, all visibilities from all pointings are stacked and gridded together onto a single tangent plane. In facet-based imaging, the Mueller matrices define a point spread function (PSF) and a PB normalisation at the centre of each facet; hence, each pointing in a mosaic will have its own set of unique Mueller matrices. It follows from Equation 20 onwards in \citet{2018A&A...611A..87T} that the apparent dirty image is:
    \begin{align}
    \widetilde{\mathbf{y}}_{\nu} & = \left( \widetilde{\mathbf{\mathcal{M}}}_{\nu}^{H} \right)^{-1} \mathbf{\mathcal{A}}_{\nu}^{H} \mathbf{\mathcal{W}}_{\nu} R_{\nu} \mathbf{v}_{\nu} \\
                            & = \left( \widetilde{\mathbf{\mathcal{M}}}_{\nu}^{H} \right)^{-1} \langle \omega_{\mathbf{b}_{\nu}} \mathbf{C}_{\mathbf{b}_{\nu}} \mathbf{\mathcal{M}}_{\mathbf{b}_{\nu}} \rangle_{\Omega_{\nu}} R_{\nu}  \mathbf{x}_{\nu}.
    \label{eq:2}
    \end{align}
    Here, $\widetilde{\mathbf{y}}_{\nu}$ is the apparent dirty image vector, $\widetilde{\mathbf{\mathcal{M}}}_{\nu}$ is the normalisation associated with the PB attenuation given by $\widetilde{\mathbf{\mathcal{M}}}_{\nu} = \sqrt{\left\langle \omega_{\mathbf{b}_{\nu}} \mathbf{\mathcal{M}}^{H}_{\mathbf{b}_{\nu}} \mathbf{\mathcal{M}}_{\mathbf{b}_{\nu}} \right\rangle_{\Omega_{\nu}}}$, where $\omega_{\mathbf{b}_{\nu}}$ are the individual visibility weights in the diagonal matrix $\mathbf{\mathcal{W}}_{\nu}$. Then,  $\mathbf{C}_{\mathbf{b}_{\nu}} = \mathbf{\mathcal{F}}^{H} \mathbf{S}^{H}_{\mathbf{b}_{\nu}} \mathbf{S}_{\mathbf{b}_{\nu}} \mathbf{\mathcal{F}}$ is a convolution, and $R_{\nu}$ is a rotation matrix of the form:
\begin{equation}
R_{\nu}(\alpha_{\mathbf{b}_{\nu}}) = \begin{bmatrix} \cos(\alpha_{\mathbf{b}_{\nu}}) & \sin(\alpha_{\mathbf{b}_{\nu}}) \\ -\sin(\alpha_{\mathbf{b}_{\nu}}) & \cos(\alpha_{\mathbf{b}_{\nu}}) \end{bmatrix}, \quad
\alpha_{\mathbf{b}_{\nu}} = 2\pi \frac{\nu}{c} \cdot \mathbf{b}_{\nu} (\mathbf{s} - \mathbf{s}_0).
\label{eq:rotation_matrix}
\end{equation}        
        This represents a 2D geometric phase shift to rotate all gridded visibilities (3D for ungridded visibilities) to the same phase centre, where $\alpha_{\mathbf{b}_{\nu}}$ is the phase shift per baseline-time-frequency sample, $c$ the speed of light, $\mathbf{s}$ the sky direction unit vector and $\mathbf{s}_0$ the unit vector in the direction of the chosen phase centre. Deconvolution proceeds as normal, and the mosaiced skymodel is predicted to each of the input datasets. 
 

        In practice, mosaicing in \textsc{DDFacet} is made very easy. For any radio interferometer with appropriate PB models, the user simply specifies the beam parameters in the \textsc{--Beam} section and the mosaic coordinate centre with the parameter \textsc{--Image-PhaseCenterRADEC}. Then, given any number of standard Measurement Sets with the same spectral window structure, \textsc{DDFacet} will grid all visibilities onto a single tangential plane, perform the phase rotation of Equation \ref{eq:rotation_matrix} to the stated centre, and apply the appropriate PB normalisation per facet, producing a facetted visibility plane mosaic. To our knowledge, only \textsc{DDFacet} currently has the capability to perform this procedure for the MeerKAT telescope and its primary beam. CASA's \textsc{tclean} can perform this operation when using its \textsc{mosaic}, \textsc{awproject}, or \textsc{aw2} gridders, but requires position-independent PSFs (i.e., no facetting) and is only validated for ALMA \citep{2009IEEEP..97.1463W} and the VLA \citep{2011ApJ...739L...1P}. \textsc{WSClean} provides the external tool \textsc{chgcentre}, which performs a similar operation. Once all observations have been rotated, its IDG can then be used to make a visibility plane mosaic if appropriate PB models are available. PB models available to \textsc{WSClean} through the EveryBeam package\footnote{\url{https://git.astron.nl/RD/EveryBeam}} \citep{ASTRON2020EveryBeam} include LOFAR, MWA \citep{2013PASA...30....7T} and VLA. Furthermore, to our knowledge, none of these mosaicing approaches, except that of \textsc{DDFacet}, can include the piecewise-constant DD solutions estimated by the Wirtinger calibration of \textsc{killMS}, which are multiplied to the beam model $\widetilde{\mathbf{\mathcal{M}}}_{\nu}$ in Equation \ref{eq:2}.
        
        \section{Simulations}

        Here we present a series of simulations that aim to quantify the benefits and costs of doing 3GC visibility plane mosaicing in \textsc{DDFacet}. We used the Measurement Sets from Table \ref{tab:Table 1} as a framework for the simulated data.
        

        \begin{table}
		\caption{Observation details of the data used in this work. It consists of two MeerKAT L-band pointings of bandwidth 856-1712 MHz, centred on 1283 MHz, with 4096 frequency channels and an 8 second dump rate.}
		\centering
		\begin{tabular}{ll}
			\hline
			Pointing 1 \\
			\hline
                RA, DEC (J2000) & 13h31m08s, -31$^{\circ}$40$^{\prime}$23$^{\prime\prime}$  \\
                On-source integration & 6 hrs \\
			Observation date & 06 July 2018 \\
            Schedule Block & 20180706-0112 \\
                \hline
			Pointing 2 \\
			\hline
                RA, DEC (J2000) & 13h33m35s, -31$^{\circ}$40$^{\prime}$30$^{\prime\prime}$  \\
                On-source integration & 3 hrs \\
			Observation date & 07 July 2019 \\
            Schedule Block & 20190522-0026 \\
			\hline
		\end{tabular}
		\label{tab:Table 1}
	\end{table}

        \begin{figure}
		\centering
            \includegraphics[width=\columnwidth]{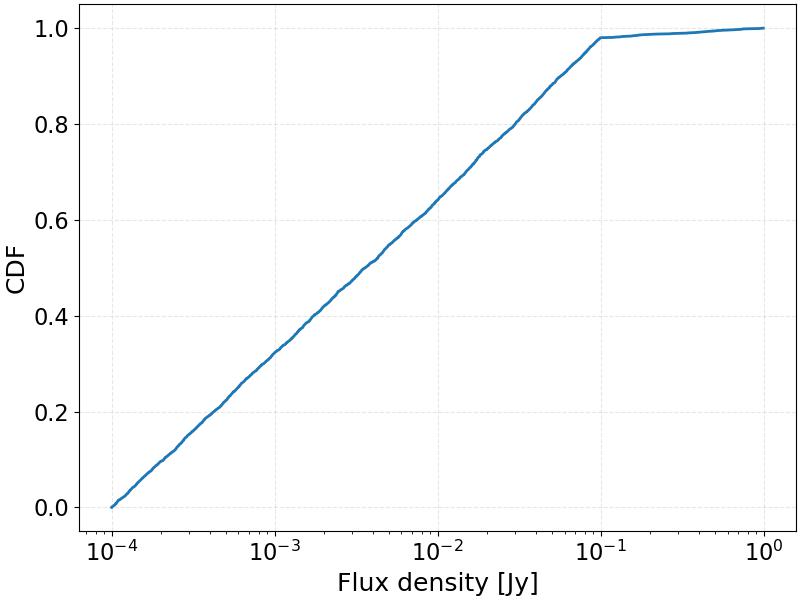}
		\caption{Source intensity cumulative density function (CDF) for the suite of simulations.} \label{fig:sim_source_intensities}
	\end{figure}

        \subsection{Setup}

        \begin{figure*}
		\centering
            \includegraphics[width=\textwidth]{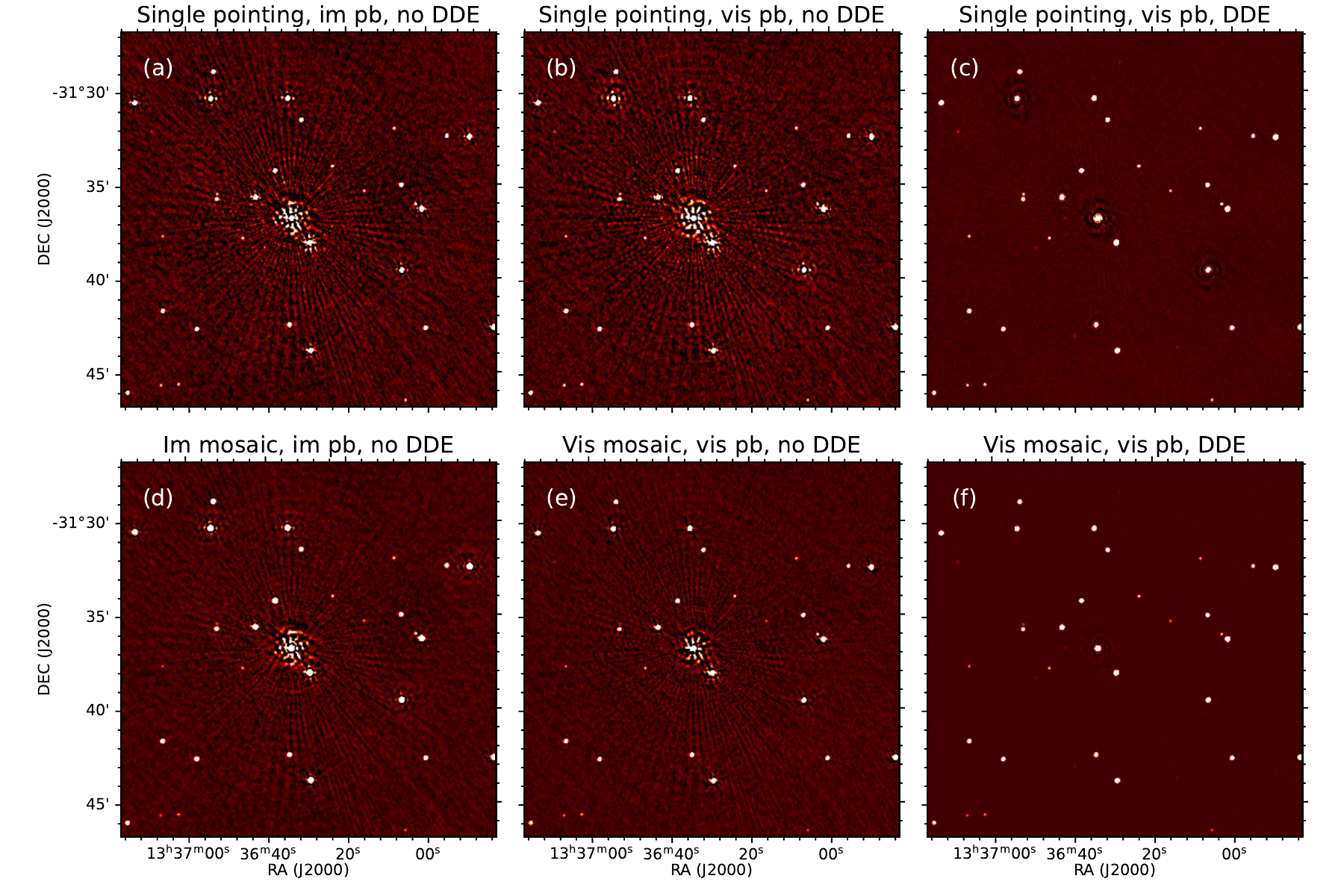}
		\caption{Cutouts of the simulated images. Abbreviations: Im = image, vis = visibility, PB = primary beam, DDE = direction-dependent effects. Root mean squared (rms) noises: 17.2, 17.0, 5.1, 10.1, 9.7, 1.9 $\mu$Jy/beam} \label{fig:sims_suite}
	\end{figure*}

        The effectiveness and reliability of any mosaicing procedure will strongly depend on the nature of the individual observations, i.e., deep and sparse, or shallow and dense, or some combination thereof. We chose two medium-deep pointings of a science target currently under investigation in a parallel work (Trehaeven et al. In Prep) to test the 3GC mosaicing. In this case, to study how well visibility plane mosaicing, direction-dependent calibration and PB correction recover source flux density and spectral index in a realistic scenario, we simulated a test population of 5000 point sources in two mock MeerKAT L-band (856 - 1712 MHz, central frequency 1283 MHz) observations, averaged to 1024 channels, separated by $\sim$0.5$^{\circ}$. The positions of these pointings are shown in the top panel of Figure \ref{fig:ssc_and_bridge}. The UV-coverage was built up with 15-minute scans for six and three hours, respectively. The confusion limit for such observations is $\sim$4.5 $\mu$Jy/beam at a resolution of $\sim7^{\prime\prime}$. The simulated components were randomly distributed throughout a $4^{\circ}\times4^{\circ}$ field of view (FoV), uniformly distributed on a logarithmic intensity scale from 0.1 mJy to 0.1 Jy with 100 sources scattered up to 1 Jy. The spectral indices were uniformly distributed on a linear scale from -3 to 3. Figure \ref{fig:sim_source_intensities} shows the distribution of the source intensities. The MeerKAT L-band half power beam width (HPBW) at its central frequency is $\sim1.1^{\circ}$ and therefore covers the central part of the simulated sky region. The manufactured skymodel was formatted into the native \textsc{DDFacet} DicoModel format and predicted into visibilities using the option \textsc{--Predict-InitDicoModel}. Random Gaussian noise, commensurate with real data, was added to simulate realistic observational conditions. These two mock pointings were imaged, PB corrected, and then mosaiced in both the image and visibility planes for comparison. An analogous procedure could be performed with, say, \textsc{WSClean} to set up a similar suite of simulations, except that by including \textsc{killMS} solutions into the predict, we have introduced significant DDEs.

        \subsection{Imaging and mosaicing}
        
        We used the SubSpace Deconvolution algorithm (SSD2) throughout the simulations. SSD2 performs joint deconvolution on localised groups of pixels (islands) in an image, using a genetic algorithm for subspace optimisation, making it more robust to local PSF variations due to DDEs. Each pointing was imaged for two SSD2 major cycles with automasking and a third-order spectral polynomial over three subbands. A 26x26 square facet grid was imposed onto the FoV within which a single PSF was evaluated along with the PB (when enabled). After the automasked major cycles, a manual mask was created with the masking tool \textsc{breizorro}\footnote{\url{https://github.com/ratt-ru/breizorro}} \citep{2023ascl.soft05009R}, which is an expanded and packaged version of the \textsc{DDFacet} utility \textsc{MakeMask.py}. The deconvolution was then continued for one more major cycle with the external mask. Each mock observation was imaged twice to compare PB correction in the image versus visibility planes, i.e., with the PB section in \textsc{DDFacet} disabled and then enabled. We used the array-averaged PB models released by \citet{2023AJ....165...78D} throughout. Image plane PB correction was performed using \textsc{spimple}\footnote{\url{https://github.com/landmanbester/spimple}}\citep{Bester2025Spimple}, which is a tool to perform image-based convolution, primary beam correction and spectral index fitting. We use its functionality to interpolate the PB models across parallactic angle and convert them to a time-averaged Stokes I power beam. The correction is applied by simply dividing the final image by this power beam. Visibility PB correction was performed during deconvolution to account for the time- and frequency-dependent beam variations in each facet during the mock observations. 3GC solutions, derived using killMS, from the set of real-world observations, described in Section \ref{sec:vis_plane_mosaic_real}, were applied to produce a 3GC single pointing image.

        \begin{figure*}
		\centering
            \includegraphics[width=\textwidth]{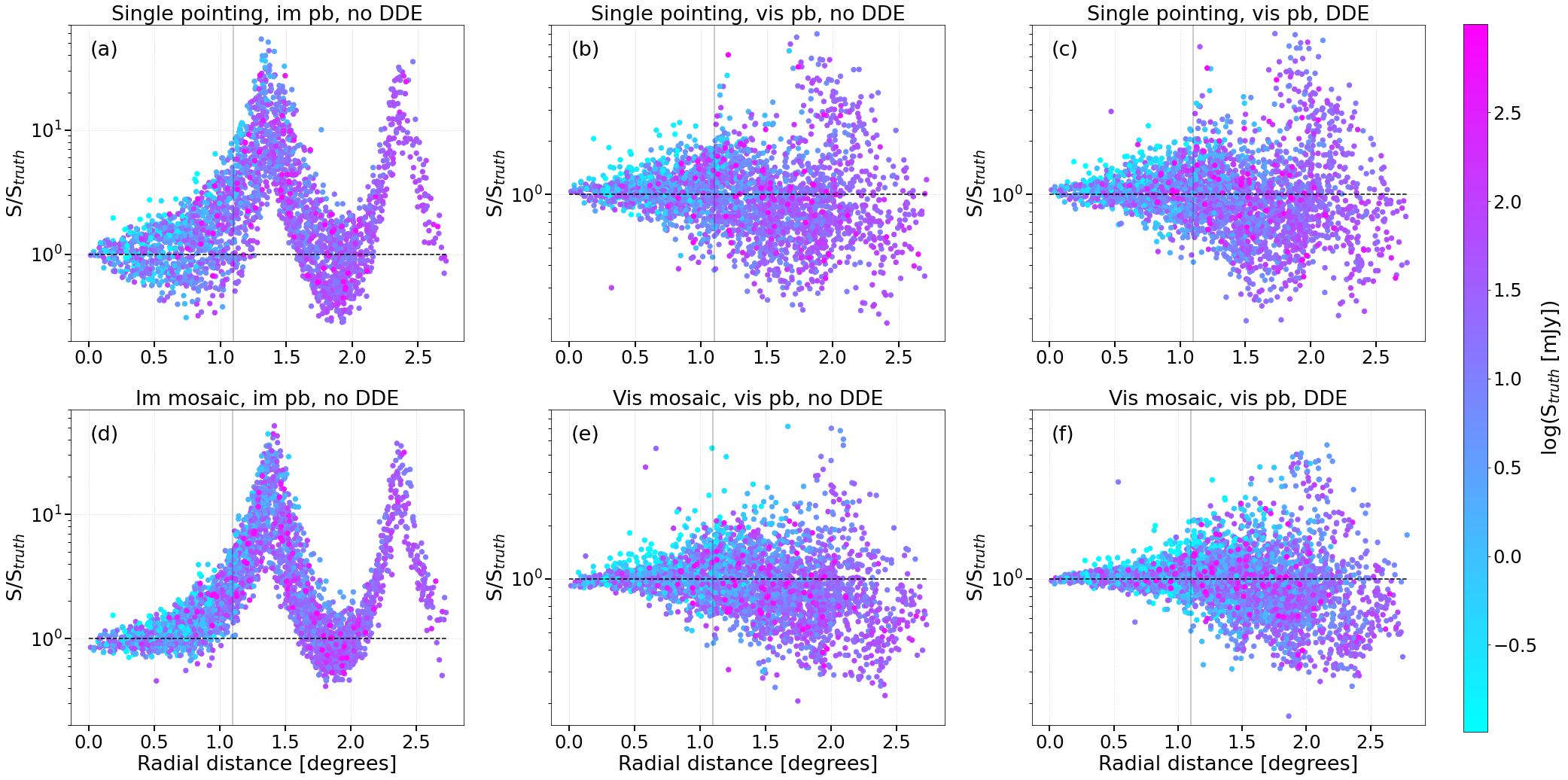}
		\caption{Recovered point source peak flux densities for the various simulations. The vertical dashed lines indicate the edge of the PB main lobe at a power of 0.1. The horizontal dashed lines show the unity level. Each simulation is labelled (a)-(f) according to the text.} \label{fig:sim_fluxes}
	\end{figure*}
    
        \begin{figure*}
		\centering
            \includegraphics[width=0.9\textwidth]{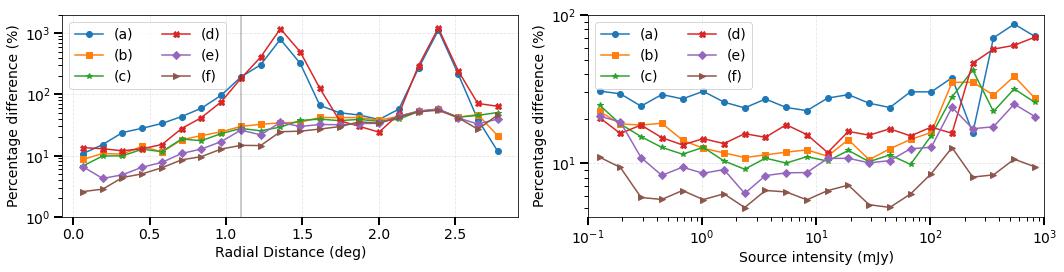}
            \includegraphics[width=0.9\textwidth]{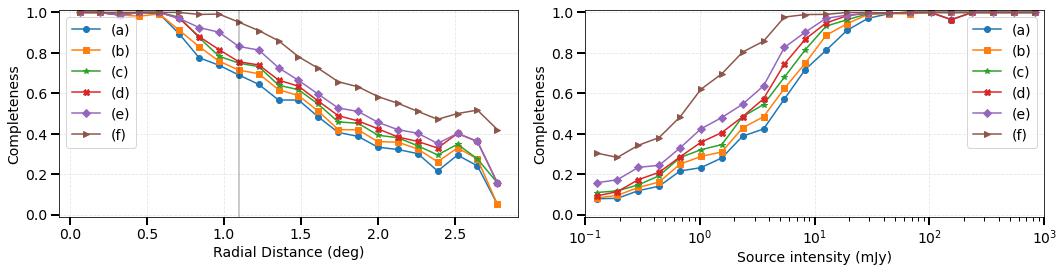}
		\caption{Average percentage difference (\textit{top}) and completeness fractions (\textit{bottom}) as a function of radial distance (\textit{left}) and source intensity (\textit{right}). Each simulation is colour-coded.} \label{fig:sim_completeness}
	\end{figure*}

        Image plane mosaicing was performed by convolving each PB-disabled image to a common circular beam size (to the nearest whole number) and using the \textsc{MONTAGE} to reproject them onto a single plane and co-add them, weighted by the square of the individual power beams. Visibility plane mosaicing was automatically performed in \textsc{DDFacet} by specifying the new phase centre with the option \textsc{--Image-PhaseCenterRADEC}, chosen to be roughly the centre of the intercluster radio bridge. The same settings and imaging logic as the single-pointings were used. Finally, similar to the single pointing, direction-dependent solutions were applied to produce a 3GC, PB corrected visibility plane mosaic. Figure \ref{fig:sims_suite} shows a cutout of each simulated image, from top left to bottom right: (a) single pointing with image plane PB correction and no direction-dependent solutions, (b) single pointing with visibility plane PB correction and no direction-dependent solutions, (c) single pointing with visibility plane PB correction with direction-dependent solutions, (d) image plane mosaic with image plane PB correction and no direction-dependent solutions, (e) visibility plane mosaic with visibility plane PB correction and no direction-dependent solutions, (f) visibility plane mosaic with visibility plane PB correction and direction-dependent solutions.

        \subsection{Simulation results}

        The purpose of these simulations is to test how well the different mosaicing, PB correction and direction-dependent calibration methods recover source flux density and spectral index. After imaging, we measure the source intensity by running PyBDSF \citep[Python Blob Detection and Source Finder,][]{2015ascl.soft02007M} on the frequency-averaged (or multi-frequency-synthesis, MFS) image of each simulation from (a)-(f) and extract the source flux and position. Positional crossmatching between the PyBDSF catalogues and the ground-truth catalogue was performed using the Astropy \textsc{match\_coordinates\_sky} function with a crossmatching radius taken to be approximately half the restoring beam size of the simulated images, or 3$^{\prime\prime}$. Figure \ref{fig:sim_fluxes} shows the fractional flux density (measured/true) of each crossmatched detection as a function of radial distance for all our simulation types. The radial zero-point in all panels was taken as the phase centre of the visibility plane mosaic. All panels accurately recover the source intensity at the phase centre. However, the image plane PB correction in panels (a) and (d) degrades quickly as we go further out into the image. Panel (d) has a tighter spread than panel (a) due to its mosaicing. Interestingly, panels (a) and (d) show the imprints of the PB sidelobe nulls around $1.4^{\circ}$ and $2.4^{\circ}$ with much higher recovered flux densities at these radii, having almost two orders of magnitude discrepancy. On the other hand, this trend is eliminated in the rest of the panels, all with visibility plane mosaicing, showing that the time and frequency dependent variations of the PB need to be corrected if accurate flux densities are to be measured beyond the phase centre. We see, from panels (b), (c), (d) and (f), that as we increase the simulation complexity, the spread in the measured flux densities becomes tighter and tighter. To quantify the improvement, we calculate the weighted percentage difference between the measured and true fluxes for all detected sources as a function of radial distance and source intensity, shown in Figure \ref{fig:sim_completeness}. For all cases, the most reliable flux densities are found near the phase centre and then decay radially, as expected. The difference is fairly constant through source intensity, but increases for the strongest sources, most likely caused by the fact that, due to the PB attenuation, only bright sources are detected at large radii where the deviation is greatest. Cases (a) and (d) display the same radial trend as their counterparts in Figure \ref{fig:sim_fluxes}. For the rest of the cases, the average percentage difference steadily decreases from $\sim$14\% to $\sim$6\% as the simulation complexity increases, as is suggested from the spread in Figure \ref{fig:sim_fluxes}.

        \begin{figure*}
		\centering
            \includegraphics[width=\textwidth]{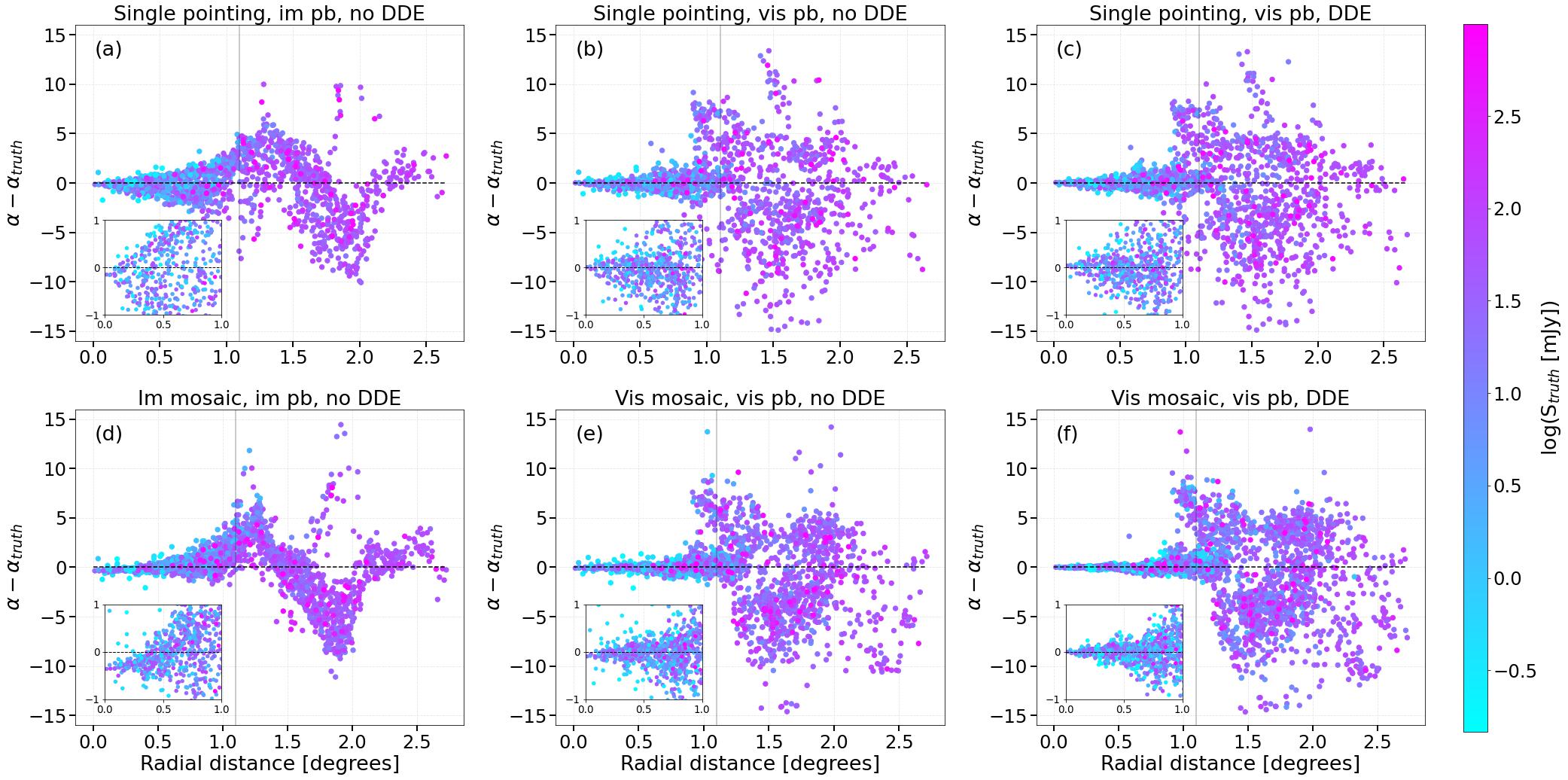}
		\caption{Recovered point source spectral index for the various simulations. The vertical dashed lines indicate the edge of the PB main lobe at a power of 0.1. The horizontal dashed lines show the zero level. The inset zooms into the inner 1$^{\circ}$ to emphasise the spread.} \label{fig:sim_spx}
	\end{figure*}

        \begin{figure*}
		\centering
            \includegraphics[width=0.9\textwidth]{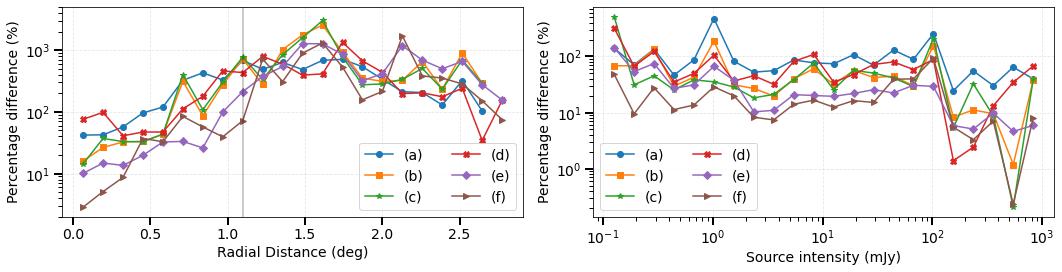}
            \includegraphics[width=0.9\textwidth]{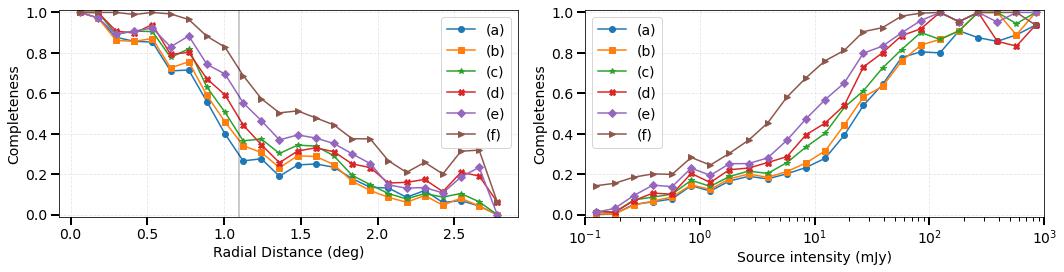}
		\caption{Average percentage difference (\textit{top}) and completeness fractions (\textit{bottom}) as a function of radial distance (\textit{left}) and source intensity (\textit{right}). Each simulation is colour-coded.} \label{fig:sim_completeness_subbands}
	\end{figure*}

        We also examine a quantity called the \textit{completeness} for each simulation setup. Completeness is defined as the fraction of true sources that are successfully recovered by the source finder, given as:
        \begin{equation}
            \text{Completeness} = \frac{\text{Tp}}{\text{Tp + Fn}} \label{eq:completeness}
        \end{equation}
        where Tp denotes the number of true positive detections and Fn denotes the number of false negative detections. Here, true positives refer to sources recovered by the source finder for which a simulated true source is located within the crossmatching radius of 3$^{\prime\prime}$ and are exactly the sources plotted in Figure \ref{fig:sim_fluxes}, and false negatives refer to true sources that the source finder was unable to detect. Since this quantity is derived from the counts of a source catalogue, it is dependent on the source finder and the particular parameters used. Nevertheless, it is instructive because it is sensitive to the noise in the image, meaning that lower noise should result in greater completeness. This quantity was calculated for each imaging setup as a function of radial distance and source intensity, shown in the bottom panels of Figure \ref{fig:sim_completeness}. The simulations are fully complete near the phase centre and for bright sources, and smoothly decrease to zero at large radii and for fainter sources. This is expected as the PB attenuation is negligible at the phase centre but becomes more and more significant further out into the image, making even the brightest sources difficult to detect. More importantly, the completeness in both radial distance and source intensity increases as we increase the imaging complexity, with the full visibility plane mosaic, case (f), being almost fully complete within 1$^{\circ}$ and above $\sim5$mJy, almost double the distance and intensity compared to its image plane counterpart (d). 

        We also examine how well the different simulations recover the intrinsic source spectral index. To this end, we ran PyBDSF on each of the three subband images for each simulation. We then cross-matched all three subband catalogues and fitted a power law to the recovered flux densities across frequency. In Figure \ref{fig:sim_spx}, we plot the resulting spectral index relative to the ground truth as a function of radial distance. All combinations of PB correction and mosaicing were insufficient to recover accurate spectral indices beyond $\sim1^{\circ}$. On the other hand, all simulations recovered accurate spectral indices at the phase centre. Thereafter, the spread increases rapidly in the image plane PB corrected images. The visibility plane treatment gives tighter results with a more gradual degradation until $\sim1^{\circ}$, progressively improving with increasing complexity. 
        
        Figure \ref{fig:sim_completeness_subbands} shows the percentage difference as a function of radial distance and source intensity for the simulated spectral indices. Similar to the flux densities, increasing the simulation complexity improves the reliability of the spectral indices, decreasing from a mean difference of $\sim$125\% down to $\sim$20\%, with the most reliable measurements being bright sources near the phase centre. We also compute the completeness for the sources detected in all three subbands. The behaviour against radial distance and source intensity is similar to the wideband case, but with a more rapid decline in completeness. Here, all cases are fully complete only near the phase centre, except for the final 3GC visibility plane mosaic, which extends out to $\sim$0.75$^{\circ}$ and then falls off similarly to the rest.
        


        \begin{figure*}
		\centering
		\includegraphics[width=0.9\textwidth]{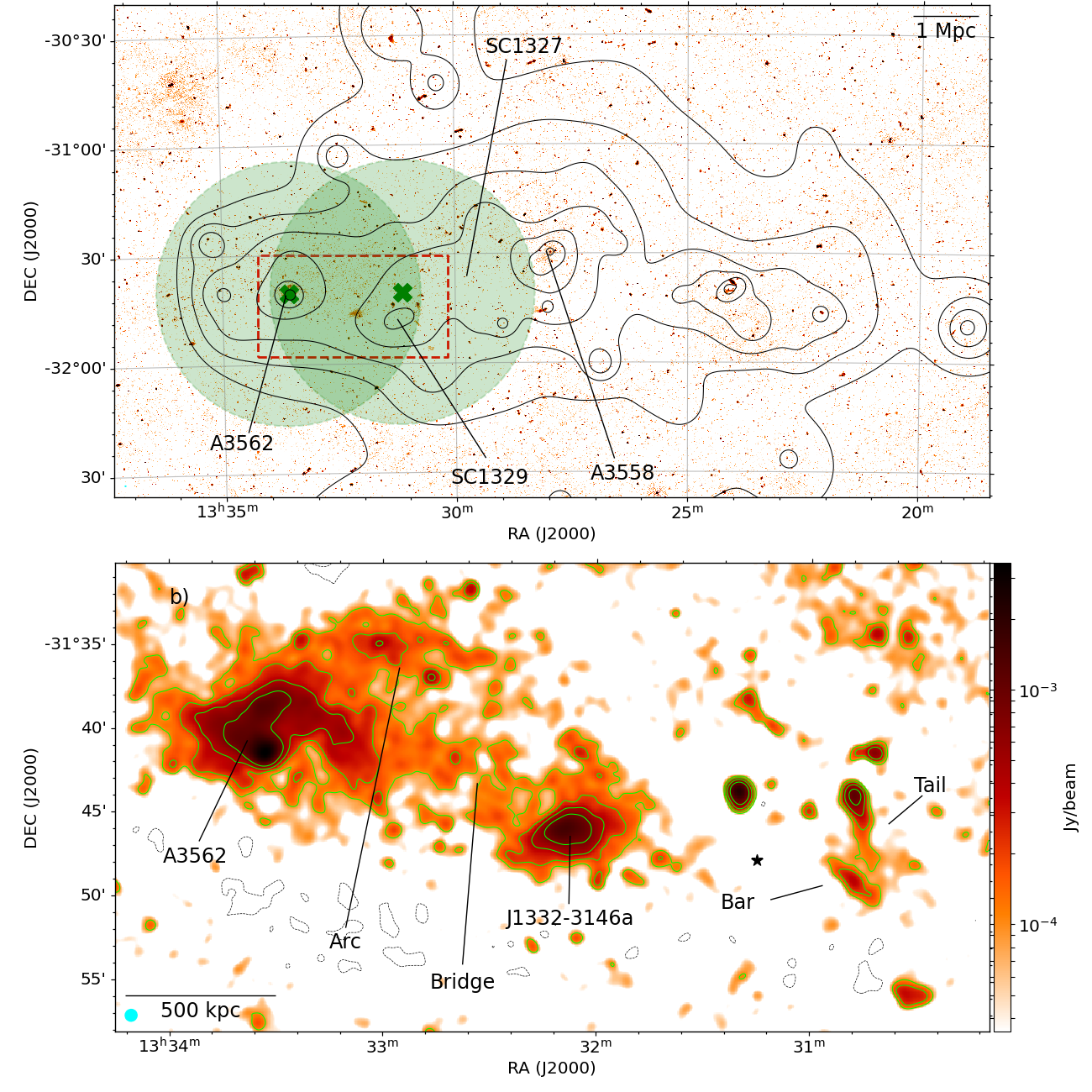}
		\caption{The Shapely Supercluster and the intercluster bridge adapted from V22 Figures 2 and 3. \textit{Top:} ASKAP 887 MHz image of the Supercluster Core with some of its member clusters and groups labelled. Black contours show the galaxy number density \citep{2018MNRAS.481.1055H}, starting from 5 galaxies/Mpc$^{2}$ and doubling thereafter. The green crosses and circles show the MeerKAT L-band pointing centres and PB HPBW. The red rectangle shows the bridge region displayed in the bottom panel. \textit{Bottom:} MeerKAT L-band 1283 MHz image of the bridge region with a $\sim$40$^{\prime\prime}$ resolution, shown in the bottom left corner, and local noise of $\sim$35 $\mu$Jy/beam. Green contours show the 3$\sigma$ level and doubling thereafter. Dotted contours show the -3$\sigma$ level. The black star shows the position of the SC\,1329--313 density peak.}
		\label{fig:ssc_and_bridge}
	\end{figure*}

        \section{Application to real MeerKAT data}
        \label{section:individual_pointings}

        In the previous Section, we used simulations to showcase the potential of visibility plane mosaicing and PB correction to improve image quality and the fidelity of recovered source flux densities and spectral indices. Here we apply these techniques to real-world data and highlight their usefulness in a practical scenario. The raw data used here can be downloaded from the SARAO archive\footnote{\url{https://archive.sarao.ac.za/}} and are summarised in Table \ref{tab:Table 1}. We recalibrate the raw data to exploit more powerful calibration software than what was available when the data was previously published by \citet{2022A&A...660A..81V}, hereafter V22. Below, we describe the test field and then the calibration and imaging of the data.


        \subsection{The Shapley Supercluster}
        
        We chose the Shapley Supercluster environment \citep{10.1093/mnras/sty2338,10.1093/mnras/staa1766} to test the visibility plane mosaicing capabilities of \textsc{DDFacet} - shown in Figure \ref{fig:ssc_and_bridge}. It is the most massive supercluster in the local Universe \citep[][M$_{\rm tot}=5\times10^{16}$ h$^{-1}$ M$_{\odot}$]{2003A&A...405..425E} at a mean redshift of $\langle z \rangle \sim 0.048$. This field is the ideal test-bed environment because of its combination of bright compact sources \citep{2018A&A...620A..25D,2025A&A...694A..28D} and large-scale faint diffuse emission \citep[V22;][]{2022ApJ...934...49G,2025MNRAS.541.2741T}, requiring accurate calibration and imaging with a high dynamic range over a vast area to exploit all the science available across a wide spatial scale and spectral range. Various galaxy clusters and groups are currently interacting within the supercluster, making it a very dynamic environment \citep{2006A&A...447..133P,2020A&A...638A..27Q}. As such, various diffuse radio sources are observed and are thought to trace the turbulence induced by the merger activity \citep[][and references therein]{2022ApJ...934...49G}. Extensive mosaiced observations of this field have been carried out with uGMRT in Band-3 \citep{2024MNRAS.533.1394M}, MeerKAT L-band (V22) and UHF-band \citep{2025MNRAS.541.2741T}. We use the archival MeerKAT L-band observations used by V22 to achieve the deepest radio image of the Supercluster core to date. The observational details are summarised in Table \ref{tab:Table 1}, and the environment is shown in Figure \ref{fig:ssc_and_bridge}. In particular, V22 used Pointing 1 to present the detection of the first GHz frequency intercluster radio bridge. This diffuse source spans $\sim$1 Mpc and has an average brightness of 0.09 $\mu$Jy/arcsec$^{2}$ at 1.28 GHz. This structure lies exactly at the intersection of the two L-band pointings. From X-ray observations, it has been suggested that the bridge traces the trajectory of group SC\,1329--313 moving east above A\,3562 to its current position, and so traces the turbulence injected into the intercluster medium during this fly-by interaction \citep{2002A&A...382...17B,2004ApJ...611..811F}. More than two decades later, we are beginning to see the faint radio signatures associated with this action. Additionally, a nearby head-tail galaxy with a bar of disjointed emission is thought to trace some ram pressure stripping or merger shock mechanism (V22). Widefield deconvolution, with the lowest possible noise, is necessary to study this highly dynamic region in detail.

        \subsection{Calibration of individual pointings}
        
        The individual pointings were calibrated independently up to and including direction-independent calibration (or second-generation calibration, 2GC). Like V22, the \textsc{CARACal} pipeline \citep{2020ascl.soft06014J} performed flagging, transfer-calibration, flux-scaling and frequency-averaging down to 1024 effective channels. Thereafter, the \textsc{Stimela2} scripting framework \citep{2025A&C....5200959S} was used to construct our own self-calibration pipeline, which allowed for greater flexibility than the current \textsc{CARACal} self-calibration workflow. All \textsc{CARACal} and \textsc{Stimela2} scripts used are made available\footnote{\url{https://github.com/ktrehaeven/Shapley}}. 
        
\begin{figure}
		\centering
            \includegraphics[width=\columnwidth]{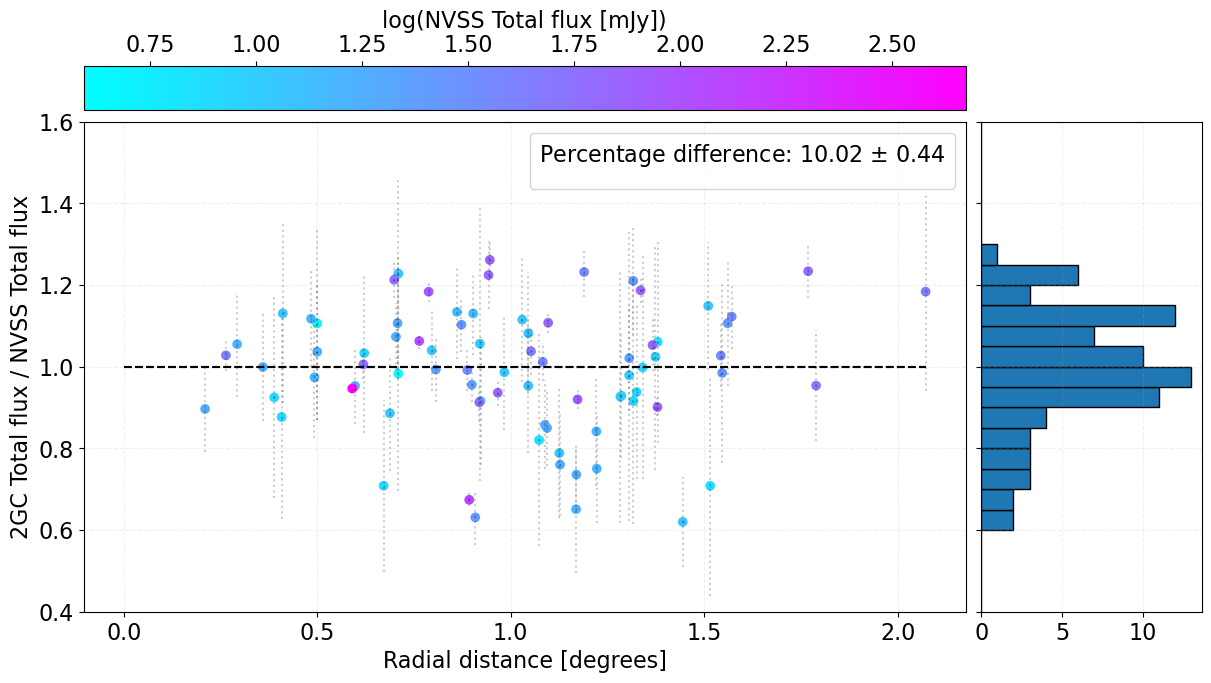}
		\caption{Point source flux density compared against NVSS as a function of radial distance, coloured by NVSS total intensity. The flux density uncertainties are plotted as light dashed grey errorbars. The horizontal dashed line shows the unity level on the y-axis. The weighted percentage error is 10.02$\pm$0.44\%.}
		\label{fig:nvss_v_2gc_fluxes}
	\end{figure}

        For 2GC, we used \textsc{DDFacet} to image and \textsc{QuartiCal} \citep{2025A&C....5200962K} to solve for direction-independent calibration solutions, but any imager/solver combination can be used. \textsc{QuartiCal} provides powerful calibration routines that exploit complex optimisation. Its most notable feature is the ability to chain various Jones terms within a single self-calibration round, allowing for more physically accurate solutions to be derived within fewer self-calibration rounds. Similar to the simulations, deconvolution was performed using the SSD2 deconvolver with a third-order spectral polynomial covering three subbands and a $3^{\circ}\times3^{\circ}$ FoV with a pixel size of 1.2$^{\prime\prime}$. A 20x20 square facet grid was imposed within which a single PSF and time- and frequency-dependent PB corrections were evaluated. We used the same PB models used in the simulations. The same imaging logic as the simulations was used, i.e., two SSD2 major cycles with native automasking, then a final major cycle with a deep manual mask generated from \textsc{breizorro}. Since we apply the beam, the skymodel is in intrinsic scale. Therefore, before calibration solutions could be derived, we were required to perform a prediction step to convert the model to the apparent scale to match the visibilities. This apparent skymodel was fed to \textsc{QuartiCal}, which iterated three times over a chain consisting of a term describing the geometric delays and associated phase shift (K, \textsc{delay\_and\_phase}) of the visibilities and a term encoding the phase and amplitude information of the parallel-hand complex gains (G, \textsc{diag\_complex}). The K-term was solved per integration over the entire bandwidth, while the G-term was solved every three minutes over each quarter of the bandwidth. The G-term solution intervals were chosen to be quite conservative to ensure no flux suppression/absorption. Finer solution intervals made no difference to the final images. Additionally, we used a UV cut of > 150 m and the \textsc{robust} solver, which implements a reweighting per solver loop based on \citet{2020MNRAS.491.1026S}. Other \textsc{QuartiCal} options like calculating corrected weights and MAD (median absolute deviation) flagging of the visibilities can be useful for more problematic fields, but were not necessary in this case. The self-calibrated data was imaged with an updated mask to visualise the improvement. To determine the accuracy of the fluxscale, we convolved the 2GC image to a 45$^{\prime\prime}$ beam, ran PyBDSF, scaled the source fluxes to 1.4 GHz (assuming $\alpha=0.7$) and compared against the corresponding NVSS flux densities \citep{1998AJ....115.1693C}. This is shown in Figure \ref{fig:nvss_v_2gc_fluxes}. We find that the fluxscale is accurate to $\sim$10\% throughout the imaged area, including the PB sidelobes.

        \begin{figure*}
		\centering
            \includegraphics[width=\textwidth]{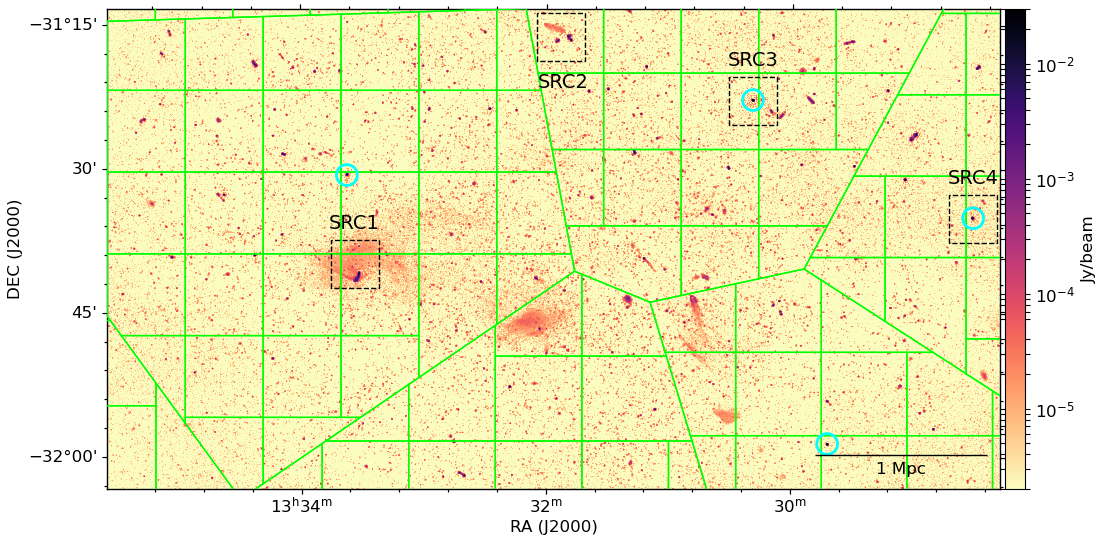}
            \includegraphics[width=\textwidth]{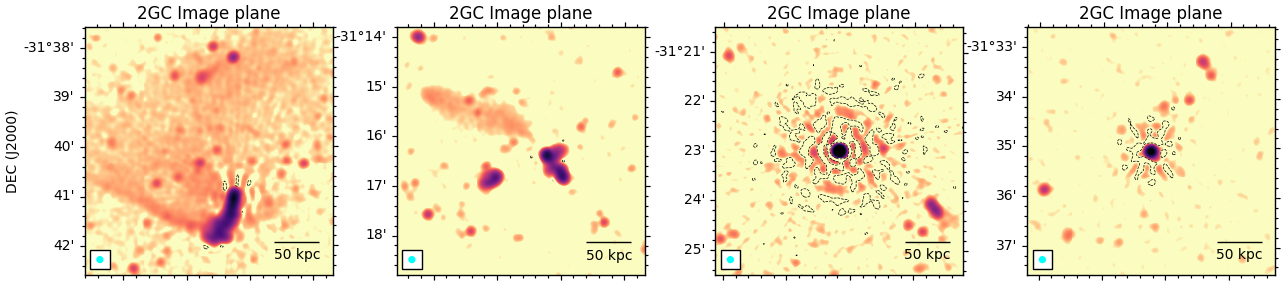}
            \includegraphics[width=\textwidth]{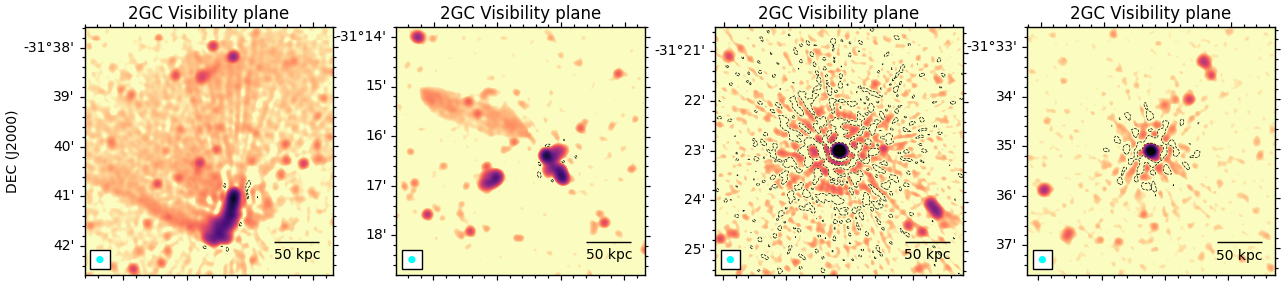}
            \includegraphics[width=\textwidth]{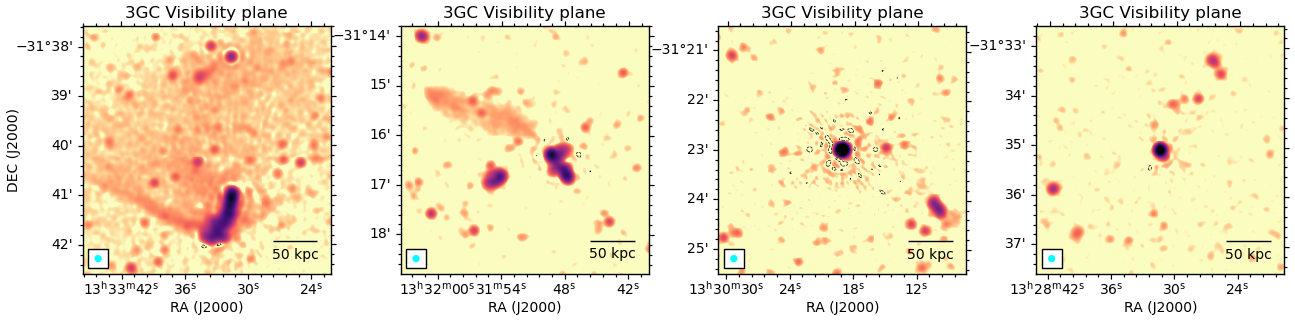}
		\caption{Final 3GC visibility plane mosaic image with examples of artefacts seen throughout calibration. All images are convolved to a 7.0$^{\prime\prime}$ circular beam for ease of comparison. \textit{Top:} the dashed black squares labelled SRC1-4 in the top panel correspond to sources shown in the grid below from left to right. The green lines show the facet and tessellation pattern. The cyan circles highlight the tessel centres. \textit{Bottom:} all dashed black contours are drawn at -18 $\mu$Jy/beam to show the level of negative artefacts. The local noise around SRC1 shown in the first column is, from top to bottom: \{11.1, 12.9, 10.8\} $\mu$Jy/beam; SRC2 in the second column: \{5.7, 5.7, 5.2\} $\mu$Jy/beam; SRC3 third column \{35.6, 44.2, 10.0\} $\mu$Jy/beam; and SRC4 in the fourth column: \{4.0, 3.6, 2.7\} $\mu$Jy/beam. The global noise within the HPBW of the PB, from top to bottom, is \{4.5, 4.8, 3.6\} $\mu$Jy/beam. All noise measurements are given in Table \ref{tab:src_noises}.}
		\label{fig:artefacts_mosaics}
	    \end{figure*}
        


        \begin{figure}
		\centering
        \includegraphics[width=\columnwidth]{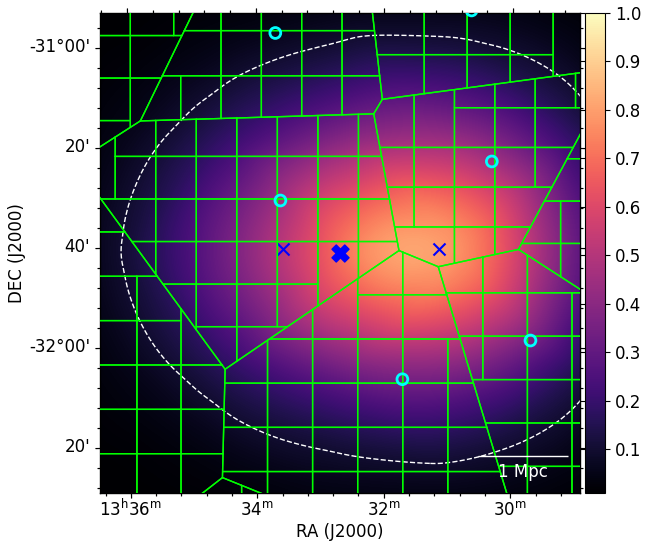}
		\caption{Example of the tessellation pattern used for 3GC overlaid on the mosaiced mean smoothed power beam. The horizontal and vertical green lines delineate facets, and the angled lines delineate the tessels. Cyan circles show the locations of the sources used to define the tessellation pattern. A small blue crosses show the individual phase centres while the bold cross on the right shows the mosaic phase centre. The dashed white line shows the 0.1 power contour.}
		\label{fig:tessellation_pattern}
	\end{figure}

        \begin{figure}
		\centering
		\includegraphics[width=\columnwidth]{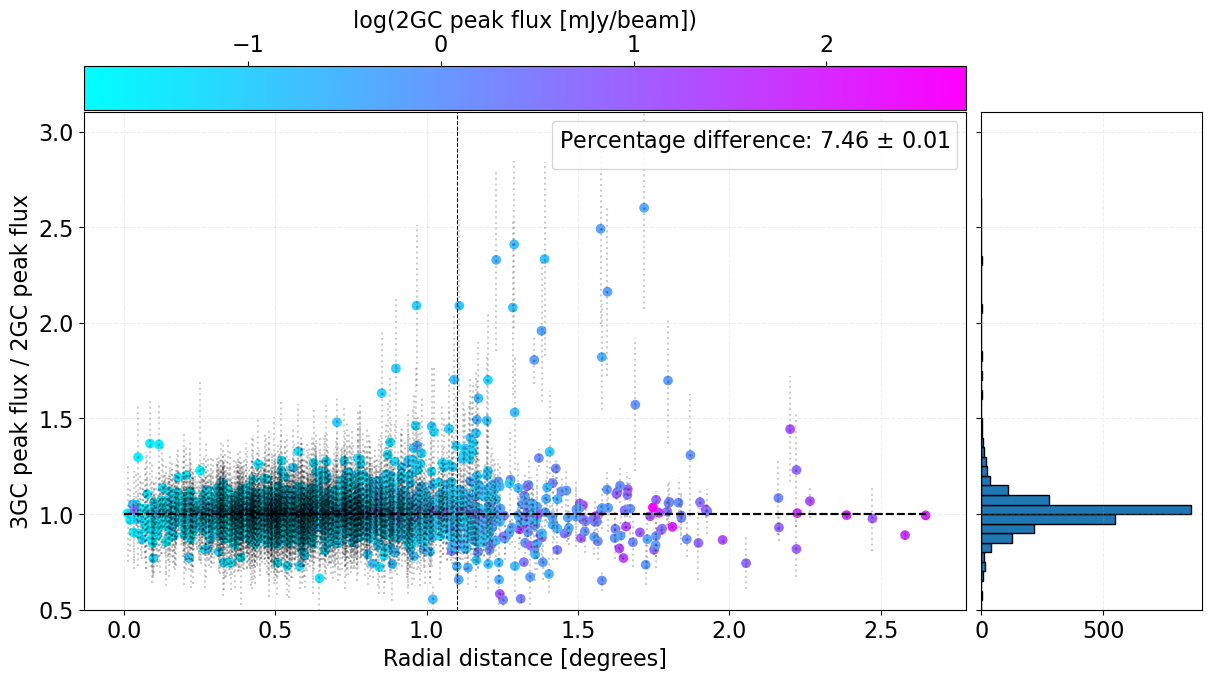}
		\caption{Point source peak flux density ratio between the 3GC and 2GC visibility plane mosaics as a function of radial distance, coloured by 2GC peak intensity. The vertical dashed line indicates the edge of the PB main lobe. The horizontal dashed line shows the unity level on the y-axis.}
		\label{fig:3gc_vs_2gc_fluxes}
	\end{figure}


        \subsection{Mosaicing}
        \label{section:mosaicing}

        We generated both image and visibility plane mosaics to compare their dynamic ranges and potential for scientific output. 

        \subsubsection{Image plane}
        
        For image plane mosaicing, similar to the simulation, we co-added the final 2GC PB corrected images using \textsc{MONTAGE}, weighted by the square of the individual beams. This is a similar procedure to the MIGHTEE and LoTSS surveys. The top row of Figure \ref{fig:artefacts_mosaics} shows a few sources around the field of view and their level of artefacts at this stage. This image is already deeper than what was presented by V22, having an average noise inside the PB of 4.5 $\mu$Jy/beam and a dynamic range (maximum/rms) of $\sim$53,000. Although some artefacts remain, this image is good enough for most use cases. However, as shown by the simulations, to get the best performance out of the available data, visibility plane mosaicing with 3GC is necessary. \textsc{DDFacet} is currently the only imager that can do this for MeerKAT data.

        \subsubsection{Visibility plane}
        \label{sec:vis_plane_mosaic_real}


        
        Before deconvolution, we first made a dirty image of the visibility plane mosaic, choosing the image centre and increasing the FoV and number of facets to match the simulations. From this dirty image, we selected the sources around which the tessellation pattern would be defined and hence the directions in which 3GC solutions would be solved for. In this case, we chose eleven of the brightest sources having residual calibration artefacts. The resulting pattern is shown in Figure \ref{fig:artefacts_mosaics} overlaid on the final 3GC mosaic, and in Figure \ref{fig:tessellation_pattern} overlaid on the mosaiced power beam. Note the elongation of the beam in Figure \ref{fig:tessellation_pattern} due to the horizontal separation of the two pointings, and the bias towards the western pointing due to its longer integration. The mosaic was imaged using the self-calibrated data and using the same logic as the simulations and individual pointings, i.e., two automasked SSD2 major cycles and then continuing with a deep manual mask for one more cycle. Examples of the 2GC visibility plane mosaic can be seen in the second row of Figure \ref{fig:artefacts_mosaics}. The 2GC image plane and visibility plane mosaics have similar fidelity, noise and artefacts. We then proceeded to 3GC, where we used the \textsc{KAFCA} solver in \textsc{killMS} to derive solutions over five-minute time intervals and 32 effective frequency channels. The solutions were then applied during a final major cycle, again continuing from the previous skymodel and with an updated mask. The bottom row of Figure \ref{fig:artefacts_mosaics} highlights the improvements. Particular improvement is shown in the bottom left panel, where radial spoke-like artefacts within the diffuse emission have now finally been corrected. Although some very minor artefacts remain around a few bright sources, the average noise within the PB improves to 3.6 $\mu$Jy/beam, constituting a 50\% improvement in dynamic range over the image plane mosaic to $\sim$80,000.
        
        It is known that 3GC often suffers from suppression/absorption, where some of the flux from bright sources or unmodelled faint sources is removed from the astrophysical signal by being partially or completely absorbed by the calibration solutions, which mistakes it for either instrumental effects or noise. Much experimentation occurred to ensure no flux suppression/absorption at the 3GC stage. Initial tests showed that a very small fraction of sources were significantly suppressed. Writing the \textsc{DDFacet} imaging weights and feeding them to \textsc{killMS} helped reduce the number of sources affected. However, we determined that a UV cut on the shortest baselines (> 150 m, similar to 2GC) eliminated this issue entirely without significantly affecting the effectiveness of the 3GC. This is the same methodology used to account for the possible biases introduced when the diffuse large-scale structure in a field has not been sufficiently modelled \citep[e.g.,][]{2016MNRAS.463.4317P}, suggesting that the presence of the intercluster bridge may be the initial cause of any suppression/absorption issues. Figure \ref{fig:3gc_vs_2gc_fluxes} shows the mean difference between the 2GC and 3GC peak fluxes of $\sim7.5\%$, which is within our fluxscale uncertainty. Additionally, a population of faint, low signal-to-noise sources tracing the outline of the main lobe of the mosaiced PB are amplified by 3GC, indicating that their PSFs may have been better constrained by the 3GC solutions. In fact, a similar behaviour is seen in the single pointings. It is possible that flux suppression can be mitigated by better modelling of diffuse emission, such as e.g., implemented by RESOLVE \citep{2016A&A...586A..76J}, SARA \citep{2012MNRAS.426.1223C}, or the large-scale shapelets approach \citep{2024A&A...692A..31Y}, but at present \textsc{DDFacet} does not implement such models.

        Although the noise significantly improves after 3GC as expected, inevitably some low-level artefacts remain around certain bright sources, such as SRC3 in Figure \ref{fig:artefacts_mosaics} and Table \ref{tab:src_noises}. Increasing the number of directions, the degrees of freedom and flexibility in the solver may improve the artefacts even further. We also note that slightly more of the large-scale diffuse emission in the Shapley bridge and arc is seen in the final mosaic than in the 2GC image plane mosaic. However, we leave the analysis of the bridge and other science targets to a later work.

        \begin{table}
        \caption{Source noises throughout the different types of mosaicing for the examples shown in Figure \ref{fig:artefacts_mosaics}. SRC1 is the leftmost example and SRC4 the rightmost, and PB represents the noise within the HPBW of each image. All images have a beam size of 7$^{\prime\prime}$, and all noise measurements have units of $\mu$Jy/beam.}
        \centering
        \begin{tabular}{|c|c|c|c|c|c|}
	\hline
	& SRC1 & SRC2 & SRC3 & SRC4 & PB\\
	\hline 
	2GC-Im & 11.1 & 5.7 & 35.6 & 4.0 & 4.5 \\
	2GC-Vis & 12.9 & 5.7 & 44.2 & 3.6 & 4.8 \\
        3GC-Vis & 10.8 & 5.2 & 10.0 & 2.7 & 3.6 \\
	\hline 
        \end{tabular}
        \label{tab:src_noises}
        \end{table}

        \section{Discussion}

        We have shown, in a fully direction-dependent framework, through simulations and the application to MeerKAT L-band data, that visibility plane mosaicing and PB correction, coupled with 3GC, improves dynamic range and completeness compared to conventional image plane mosaicing. Moreover, it provides more accurate reconstructions of source flux densities and spectral indices out to larger radii and down to lower brightness, thereby improving reliability and image fidelity. 
        
        \subsection{Potential use cases for visibility plane mosaicing}
        
        The above aspects have important implications for a number of science cases that depend on precise flux measurements over a wide FoV and bandwidth. For example, the detection and characterisation of the faint radio signatures of merger activity in the intracluster medium, such as halos, relics and intercluster bridges like those shown in Figure \ref{fig:ssc_and_bridge}. The flux density and spectral index of these diffuse low surface brightness sources are direct probes of spectral ageing and particle (re)-acceleration mechanisms \citep{2014IJMPD..2330007B,2019SSRv..215...16V}, and are used to infer the mechanics of large-scale structure formation in the Universe.

        However, the most impactful use case is for large survey projects like MIGHTEE, EMU \citep[Evolutionary Map of the Universe,][]{2021PASA...38...46N} and RACS \citep[Rapid ASKAP Continuum Survey][]{2020PASA...37...48M}, and others which aim for high completeness and uniformity across large areas. The improved image reconstruction as a function of radius shown by our visibility plane mosaics suggests that the individual pointings of an observed mosaic could be separated by greater distances while still maintaining the fidelity of the final mosaic. Thus, if data processing is planned accordingly, fewer pointings would be required, decreasing data rates and increasing survey speed. Currently, the general practice for mosaicing with the MeerKAT telescope is to separate the pointings by a distance equal to the PB FWHM/sqrt(2) at the highest frequency in the observing band. For L-band, this corresponds to a distance of $\sim0.6^{\circ}$. If visibility plane mosaicing is utilised, this metric could be increased to the entire FWHM at the highest frequency, or slightly larger. This will be particularly important for future S-band surveys due to the smaller PB at higher frequencies, and thus the need for more pointings, such as the potential southern sky polarisation survey for the MeerKAT Extension\footnote{\url{https://www.meerkatplus.tel/documents/}} (MeerKAT+). This project would observe an area $>$ 7,000 deg$^{2}$ over 3,000 hours and would produce more than 1 PB of data.

        Another interesting use case is for rapid on-the-fly (OTF) mapping projects like VLASS \citep[Very Large Array Sky Survey,][]{2020PASP..132c5001L} and MeerKLASS \citep[MeerKAT Large Area Synoptic Survey,][]{2016mks..confE..32S,2025MNRAS.537.3632M}. OTF imaging is when the telescope scans the sky repeatedly at a swift slewing speed with antenna voltages read out during a series of short snapshot measurements. Such an observing strategy allows for a large area to be surveyed very quickly. VLASS implements visibility plane mosaicing for the VLA in \textsc{CASA tclean}. MeerKLASS uses MeerKAT in single dish mode to perform auto-correlation HI intensity mapping \citep{2021MNRAS.505.3698W} and measure cross-correlation power spectra \citep{2023MNRAS.518.6262C}. It slews the antennas at $\sim5^{\prime}$ per second with an integration time of two seconds. In these cases, for imaging purposes, visibility plane mosaicing is not just advantageous but a necessity. This is because the limited signal-to-noise of a single integration may not allow for any kind of robust self-calibration, but the deeper skymodel of the mosaic, with the accumulated integration, would better constrain the calibration solutions. An OTF visibility plane mosaic could also aid in blind searches for radio transients if a particular source produces unexpected calibration artefacts or is smeared or changes throughout the snapshots with respect to the mosaic. 

        \subsection{Image vs visibility plane primary beam correction}

        We find that visibility plane PB correction provides more accurate flux densities and spectral indices than image plane PB correction beyond the half power point of the beam (and up to the 0.2 power level in the case of spectral indices). MeerKAT’s PB is not circularly symmetric, resulting in a beam gain that varies with azimuthal angle by $\sim$5\% at the nominal HPWB. Since we used the same array-averaged PB models in both cases, the overestimation of flux densities seen around the PB nulls in panels (a) and (d) of Figures \ref{fig:sim_fluxes} is the direct result of the inadequacy of the image plane implementation to account for beam squint and rotation. \citet{2022MNRAS.509.2150H} found a similar result for MIGHTEE when using the \textsc{EIDOS} PB models \citep{2021MNRAS.502.2970A}, with essentially no difference in flux density reconstruction between the image and visibility plane methods within the half power point. However, beyond this, a slight bias towards greater image plane fluxes was seen up until their radial limit of 0.7$^{\circ}$, which is what we see as well. 

        The use of the array-averaged models themselves is sufficient for this study, but for future MeerKAT+ and SKA-Mid surveys with a heterogeneous array, antenna-dependent models and details on pointing errors will be necessary to adequately correct for the beam. In this regime, detailed simulations and holographic/electromagnetic \citep{2021MNRAS.502.2970A,2023AJ....165...78D} observations would be necessary to determine if an average image plane PB correction yields any meaningful fluxes.   
       
        \subsection{Compute time}

        We have shown the proficiency of 3GC visibility plane mosaicing to produce deep radio images with high dynamic ranges. However, the fact remains that to mosaic in this manner, it is necessary to grid/degrid all visibilities from all Measurement Sets (MSs). Thus, for large datasets, imaging can be very computationally expensive. In our case, using the Rhodes University Centre for Radio Astronomy Techniques \& Technologies (RATT) compute cluster, with 32 out of a total of 64 CPUs (AMD EPYC 7702P 64-Core Processor) and 500 Gb available RAM, 2GC imaging and calibration took $\sim$38-40 hours per pointing. Most of this time went into the SSD2 genetic minor cycle deconvolution, and would depend on the chosen deconvolver. More traditional algorithms like Hogbom may be faster or slower. Within this time, the actual self-calibration with \textsc{QuartiCal} only took 2.5 and 1.5 hours per pointing. The 2GC visibility plane mosaic took 58 hours to deconvolve - a similar duration to the sum of the two individual pointings. Thereafter, 3GC with \textsc{killMS} took 74 and 18 hours, respectively. The final 3GC mosaic took an additional 65 hours for a total compute time of 216 hours for the mosaic. This is 24 times longer than the on-source integration time. For comparison, we also performed 3GC on the individual pointings, and the sum of the compute times was comparable to that of the mosaic. The final numbers will depend on the specifications and availability of the compute resources used and the particular science case. Nevertheless, the mosaic is no more expensive than the sum of the individual pointings.

        Possibilities to improve compute efficiency include the use of GPUs to accelerate processing, skipping the imaging and direction-independent calibration of the individual pointings to proceed directly to the deconvolution of the mosaic, experimenting with other deconvolvers and the use of fewer facets and degridding bands. Current efforts to optimise the \textsc{DDFacet} code base include the use of Dask to access the MSs and MPI (Message Passing Interface) for enhanced parallelisation. A non-trivial amount of time was spent manually deciding which sources best defined an appropriate tessellation pattern. To improve automation, dynamic tessellation schemes or machine-learned models could be developed to optimise facet/tessel placement and solution intervals in complex mosaics. All other steps outlined in Section \ref{section:individual_pointings} can be readily automated into next-generation pipelines using \textsc{Stimela2} \citep{2025A&C....5200959S}

    


	\section{Summary and Conclusions}
	\label{section:summary}

        In this work, we have demonstrated the proficiency of 3GC-enabled visibility plane mosaicing to produce deep, high-fidelity radio interferometric images. By integrating direction dependent calibration, visibility based primary beam correction, and joint deconvolution in the visibility domain, we have showcased and validated an alternative method to traditional image plane mosaicing. Using these techniques, we have produced the first 3GC visibility plane mosaic of the Shapley Supercluster core using MeerKAT L-band data.

        Through controlled simulations, we showed that visibility plane mosaicing significantly improves the accuracy of flux density and spectral index recovery, especially beyond the half power beam width. Our best-case 3GC mosaic produced precise flux values within a 6\% uncertainty and spectral indices within 20\% throughout the imaged area, including primary beam sidelobes, and achieves high completeness (>90\%) for sources out to twice the radii and down to half the brightness than image plane equivalents. Applying the method to MeerKAT L-band data of the Shapley Supercluster, we achieved a 20\% increase in dynamic range over the image plane mosaic. Direction-dependent artefacts were substantially reduced, and the fluxscale was accurate to within 10\% across the field of view. 
        
        This work lays a critical foundation for widefield, high-dynamic-range imaging with MeerKAT and future instruments such as SKA-MID. By enabling wider separation between pointings and recovering fainter emission with high fidelity, visibility plane mosaicing could reduce data rates and accelerate deep survey mapping and improve detection of low surface brightness sources critical to cosmic structure formation studies and many other areas of image domain astrophysics. In a follow-up paper, we will perform visibility plane point source subtraction on the MeerKAT 3GC mosaic produced in this work, alongside new UHF-band data using the same techniques, enabling a broadband analysis of the intercluster bridge in the Shapley Supercluster core. We will also extend the present Stokes I results to Stokes QUV imaging to characterise the potential of visibility plane mosaicing in full polarisation.   
	
	\section*{Acknowledgements}

	The MeerKAT telescope is operated by the South African Radio Astronomy Observatory, which is a facility of the National Research Foundation, an agency of the Department of Science and Innovation. The financial assistance of the South African Radio Astronomy Observatory (SARAO) towards this research is hereby acknowledged (www.sarao.ac.za). KT acknowledges financial support from the South African Department of Science and Innovation's National Research Foundation under the ISARP RADIOMAP Joint Research Scheme (DSI-NRF Grant Number 150551). KT and TV acknowledge partial support from the INAF mini-grant 2022 ShaSEE (Shapley Supercluster Exploitation and Exploration). OMS’s research is supported by the South African Research Chairs Initiative of the Department of Science and Technology and National Research Foundation (grant No. 81737).
    
	\section*{Data Availability}

        The raw data underlying this article is available via the SARAO archive at \url{https://archive.sarao.ac.za} under proposal ID SSV-20180624-FC-01 (Schedule Block: 20180706-0112) and SCI-20190418-OS-01 (Schedule Block: 20190522-0026). All data reduction scripts are available at \url{https://github.com/ktrehaeven/Shapley}.
        
	
	
	\bibliographystyle{mnras}
	\bibliography{main}

	
	
	\appendix


	\bsp	
	\label{lastpage}
\end{document}